\documentclass{iau} 
\usepackage{graphicx}
\usepackage{makeidx}
\usepackage{blindtext}
\usepackage{hyperref}

\title[~~Inclusive education and research] 
{Development of astronomy research and education in Africa and Ethiopia}

\author[Mirjana Povi\'c]   
{Mirjana Povi\'c$^{1,2}$}

\affiliation{
$^1$Ethiopian Space Science and Technology Institute, Ethiopia.\\ 
$^2$Instituto de Astrof\'isica de Andaluc\'ia (CSIC), Spain.\\
email: {\tt mpovic@iaa.es}.   \\[\affilskip]
	}

\pubyear{2020}
\volume{367}  
\jname{Education and Heritage in the Era of Big Data in Astronomy}
\editors{R.M. Ros, B. Garcia, S. Gullberg, J. Moldon \& P. Rojo, eds.}
\begin{document}

\maketitle

\begin{abstract}
Africa has amazing potential due to natural (such as dark sky) and human resources for scientific research in astronomy and space science. At the same time, the continent is still facing many difficulties, and its countries are now recognising the importance of astronomy, space science and satellite technologies for improving some of their principal socio-economic challenges. The development of astronomy in Africa (including Ethiopia) has grown significantly over the past few years, and never before it was more possible to use astronomy for education, outreach, and development as it is now. However, much still remains to be done. This paper will summarise the recent developments in astronomy research and education in Africa and Ethiopia and will focus on how working together on the development of science and education can we fight poverty in the long term and increase our possibilities of attaining the United Nations Sustainable Development Goals in future for benefit of all. 

\end{abstract}

\firstsection 

\vspace{-0.5cm}

\section{Introduction}
Astronomy can be used as an important tool for development and for achieving the United Nations (UN) Sustainable Development Goals (SDGs)\footnote{https://sdgs.un.org/}. Being one of the most multidisciplinary sciences\footnote{https://www.iau.org/static/education/strategicplan\_2010-2020.pdf}, it can be used efficiently to promote education and to inspire-for and promote science, contributing directly to SDGs 4 and 8 (e.g., Govender, 2009; Dalgleish, 2020). The Network for Astronomy School Education (NASE)\footnote{http://sac.csic.es/astrosecundaria/en/Presentacion.php}, NUCLIO\footnote{https://nuclio.org/}, Galileo Teacher Training Program\footnote{http://galileoteachers.org/}, and Universe Awareness (UNAWE)\footnote{https://www.unawe.org/} are only some of the examples where through astronomy tens of thousands of children, students, and teachers have been involved in science. Astronomy can also effectively contribute to socio-economical growth of countries and therefore to SDGs 8 and 10 (e.g., \cite{McBride2018}). We have great examples such as South Africa, Chile, or Canary Islands in Spain and Hawaii in US, where governments selected astronomy and space science to be priority fields for their socio-economic development (e.g., see South African Ten-Year Plan for Science and Technology\footnote{https://www.gov.za/documents/ten-year-plan-science-and-technology\#}, or the astronomy program of the Government of Chile\footnote{https://www.conicyt.cl/astronomia/files/2013/11/Roadmap\_Astronomia\_v3.pdf}). Astronomy is also one of the most cutting-edge sciences and one of the principal contributers to technological development and innovation over the last decades (e.g., Vigroux, 2009), contributing directly to different SDGs including SDG 9. Astronomy also showed over the past decades to be an important tool for promoting peace and diplomacy through large international collaborations. The Square Kilometer Area (SKA) is a great example in this aspect, and direct contributer to SDGs 16 and 17. In the digital revolution that we are leaving now we can see direct contributions of astronomy, through wifi discovery, development of computing (e.g., through grid computing), communication (e.g., satellite communication), global positioning system (GPS; e.g., through atomic clocks based on distant quasars observations), and imaging (e.g., through charged coupled devices (CCDs)), and in the future through big data and related technologies (e.g., through SKA revolution) (e.g., \cite{Rosenberg2013}). 
Benefits of astronomy for socio-economical development in Africa have been recognised by the African Union (AU) and highlighted in the Common African Position on the Post-2015 Development Agenda\footnote{https://sustainabledevelopment.un.org/content/documents/1329africaposition.pdf}. As a result of that the very first African Space Strategy has been developed under the AU, recognising space science and astronomy as important for achieving SDGs. In addition, many African countries are putting important efforts in developing astronomy and space science in terms of research, institutional development, infrastructure, human capacity development (HCD), and policies (e.g., \cite{Povic2018}). 

\vspace{-0.5cm}

\section{Astronomy developments in Africa}

Over the past years significant improvements have been done regarding astronomy development across Africa. Development of radio astronomy is one of the principal priorities through the SKA and African Very Long Baseline Interferometry Network (AVN). Through these big international collaborations, beside South Africa, different African countries are involved including: Botswana, Ghana, Kenya, Madagascar, Mauritius, Mozambique, Namibia, and Zambia. All of these countries signed for the first time in 2017 the memorandum of understanding to collaborate on development in radio astronomy. Ghana was the first country among all African SKA partners to build the radio observatory in Kuntunse, by converting the 32m telecommunication antenna into radio telescope (e.g., \cite{Wild2017}). Ghana became the third African country with professional radio astronomy facilities after South Africa and Mauritius. Starting from 2025, during the SKA phase 2, thousands of dishes are planned to be built in South Africa and other African partner countries. Recently, South African Radio Astronomy Observatory (SARAO) has been established by joining the South African SKA and HartRAO. The first phase of MeerKAT has been completed in 2018 with establishment of 64 radio array dishes of 13.5m each (e.g., \cite{Camilo2018}). Thanks to this, we are now able to obtain from Africa some of the most detailed radio images of the Universe. Mauritius, in collaboration with India, is operating a radio telescope since 1992. Recently, Nigeria successfully assembled and installed 3m radio telescope mainly for HCD purposes (e.g., \cite{Povic2018}). Namibia is now planning to build the first millimeter-wave radio telescope in Africa together with Netherlands (e.g., \cite{Backes2017}). Hydrogen Intensity and Real-time Analysis eXperiment (HIRAX) is under development in South Africa with plans to build 1000 6m dishes that will operate in 400-800 MHz window (e.g., \cite{Newburgh2016}). 

Besides radio astronomy, development in optical astronomy also experienced a lot of progress over the past years. In South Africa the world largest 11m SALT telescope (e.g., \cite{Buckley2006}) plus some another 20 South African and international telescopes are operating under the South African Astronomical Observatory (SAAO). New MeerLICHT robotic 0.65m optical telescope has been recently built in South Africa to be synchronized with MeerKAT and its radio observations. In Morocco, Oukaimeden Observatory experienced strong development since its inauguration in 2007, including the establishment of 60cm TRAPPIST-North telescope dedicated to the search of extrasolar planets (e.g., \cite{Benkhaldoun2018}). Ethiopia built in 2014 Entoto Observatory with two twin 1m telescopes (e.g., \cite{Tessema2012}). 1m telescope  has been moved from La Silla in Chile to Burkina Faso who is now trying to build its first astronomical observatory (e.g., \cite{Carignan2012}). Different countries are now working on site testing for development of new optical observatories, including Algeria, Egypt, Ethiopia, Kenya, and Tanzania (e.g., \cite{Povic2018}). Regarding development of gamma-ray astronomy in Africa, H.E.S.S observatory in Namibia with its five Cherenkov telescopes is still one of the best ground-based facilities in the field of high-energy astronomy. 

Regarding HCD in astronomy, remarkable progress has been made over the past years. New post-graduate programs have been established across the continent (e.g., in Egypt, Ethiopia, Kenya, Morocco, Namibia, Nigeria, Rwanda, South Africa, Sudan, Uganda, etc.). Office of Astronomy Development (OAD-IAU) with its constant support contributed significantly to HCD. Hundreds of African MSc students have been trained through the National Astrophysics and Space Science Programme (NASSP)\footnote{http://www.nassp.uct.ac.za/} in South Africa. Many other students benefited through trainings and their studies supported by the AVN programme, Development in Africa with Radio Astronomy (DARA; e.g., \cite{Hoare2018})\footnote{https://www.dara-project.org/}, SKA-HCD programme, International Science Program (ISP)\footnote{https://www.isp.uu.se/}, or The African Initiative for Planetary and Space Science (AFIPS)\footnote{https://africapss.org/}. Public awareness has been improved across the whole continent. In many countries primary and secondary school education benefited through different teachers training programs such as NASE, Galileo Teachers Training, or NUCLIO. The African Astronomical Society (AfAS)\footnote{https://www.africanastronomicalsociety.org/} has been established recently in 2019, with aim to be the voice of astronomy in Africa and to contribute to address the challenges faced by the continent through the promotion and advancement of astronomy, and through the activities under some of its main committees such as Science Committee, Outreach Committee, and recently established African Network of Women in Astronomy (AfNWA). Finally, the recently launched African Strategy for Fundamental and Applied Physics (ASFAP)\footnote{https://africanphysicsstrategy.org/} aims to strengthen in future the development of physics, including astronomy, across the continent.  

\vspace{-0.5cm}

\section{Astronomy developments in Ethiopia}

The Ethiopian Space Science Society (ESSS)\footnote{https://www.ethiosss.org/} was established in 2004 as a civic society whose aim was to promote the development of astronomy and space science in Ethiopia and its benefits for the society (e.g., \cite{Tessema2012}). ESSS activities were fundamental for later development of Entoto Observatory in 2014 and the Ethiopian Space Science and Technology Institute (ESSTI)\footnote{https://etssti.org/} in 2016, the first research center of such kind in Ethiopia and all East-African region. Under the ESSTI Astronomy and Astrophysics (A\&A) Research and Development Department different activities have been carried out over the past years. These include:
\begin{itemize}
 \item Running of MSc and PhD post-graduate program in A\&A, forming some of the very first MSc and PhD students in the country in the field. Currently, 5 PhD and 11 MSc students graduated under the department, and another 3 PhD and 4 MSc students are about to graduate in 2021.
 \item HCD of young staff members has been an important task since ESSTI establishment, through organisation of different trainings, workshops, seminars, etc. 
 \item Research in astronomy is one of the main activities of department, with currently three research groups being established in extragalactic astronomy, stellar astronomy, and cultural astronomy, and with a small cosmology group being under the development. Research contributed to astronomy and science developments in Ethiopia, through some of the very first publications, new international collaborations, and visibility given to the ESSTI and Ethiopia (e.g., Povi\'c, 2019).   
 \item A lot of work has been done over the past years on the institutional development of the ESSTI, starting from scratch with development of all its structure, departments, different guidelines, establishment of different committees, etc.
 \item Significant work has been done in organising international meetings, including the IAU symposium 356 on 'Nuclear Activity in Galaxies Across Cosmic Time' in 2019 that was the third IAU symposium organised in Africa in the last 100 years, 2017 IAU Middle-East and Africa Regional Meeting, 2017 IAU International School of Young Astronomers, and 8th African Space Leadership Congress in 2019. These events were fundamental for strengthening international collaborations, research, and HCD.  
 \item Significant work has been also carried out over the past years in putting the two Entoto telescopes in an operational mode, and supporting the site testings at the north of Ethiopia, close to Lalibela. 
 \item Department took also an active participation in development of some of the very first policies and road maps, such as the  Ethiopian Space Science Policy (green and white papers), and the Ethiopian Space Science and Technology Road Map (in progress).
 \item Close collaboration has been established with different schools and universities, and continues outreach programmes have been carried out. The department also carried out different activities with school teachers, including two NASE workshops.
 \item Finally, through the initiative STEM for GIRLS in Ethiopia and the ESSTI Gender Office, different activities have been carried out over the last 2 years for promoting STEM fields among secondary school girls and their teachers. \\
 \end{itemize}
 \vspace{0.05cm}
 International collaborations and knowledge transfer are fundamental in supporting the further developments in astronomy in Africa and Ethiopia for the benefit of our all society.

\end{document}